# SIMILARITY NETWORK FOR SEMANTIC WEB SERVICES SUBSTITUTION

**Chantal Cherifi**

LE2I Laboratory, Burgundy University, France
chantal.cherifi@iut-dijon.u-bourgogne.fr

**Abstract**

Web services substitution is one of the most challenging tasks for automating the composition process of multiple Web services. It aims to improve performances and to deal efficiently with Web services failures. Many existing solutions have approached the problem through classification of substitutable Web services. To go a step further, we propose in this paper a network based approach where nodes are Web services operations and links join similar operations. Four similarity measures based on the comparison of input and output parameters values of Web services operations are presented. A comparative evaluation of the topological structure of the corresponding networks is performed on a benchmark of semantically annotated Web services. Results show that this approach allows a more detailed analysis of substitutable Web services.

## 1 INTRODUCTION

From the early 2000 until nowadays, Web services have continuously gaining popularity with providers, business partners and clients because of the many advantages of the concept. The basic principles give the system flexibility and high availability. These platform-independent modular units of application logic accessible via standard Web protocols are also very attractive for their ability to be composed into more complex and more valuable services. The ultimate goal is to build new applications with little or no direct human intervention using sufficiently rich, machine-readable descriptions of Web services. In this scenario, applications are no longer written manually but assembled from a set of Web services that are available in the distributed environment of the Internet.

As a growing number of Web services are available on the Web and in organizations, finding and composing the right set of Web services is a very complex issue. Indeed, Web services are being developed independently by providers without any centralized coordination. Furthermore, they are subject to changes, relocations or even suppression and prone to failures or attacks. This can lead to temporally or definitive unavailability. In such a volatile and dynamic environment, composition is a highly complex task. A great deal of work on service architecture and semantic Web has been devoted to address the problem of Web service discovery. Discovery deals with finding a set of services that corresponds to a predetermined user request. However, after a composite service has been deployed, one or more constituents of the composite service may become unavailable. Hence, there arises a need to replace such components with other components while maintaining the overall functionality of the composite service. In order to avoid this drawback, Web service substitution intends to replace failing Web services.

Web service substitution is one of the most advanced tasks within the composition life cycle. It already has triggered a large amount of research. Work in this area mainly focuses on functional and non-functional properties, like Quality of Service (QoS), to classify substituable Web services. In this paper, we concentrate on the functional properties of Web services. In order to state that a Web service may replace another one, the Web services descriptions must be analyzed. Information on functional properties can be found in the textual description or in the service name. But the most probative and direct information is nested within the interfaces that contain operations name, and parameters name and type. In [1], Web services are organized into communities of substitutable services. Each community is associated with a specific functionality represented by an ontological concept. A functionality is materialized by a set of operations. Hence, Web services within a community meet the same need and are defined as functionally similar. In [2], similar Web services are grouped into clusters. Parameters and operations names are associated to an ontological concept which is processed by a lexical similarity measure. In [3], two degrees of interface similarity called equivalent and replacing are defined,. Equivalent Web services have the same number of operations and parameters, and parameters are of the same type. A replacing Web service has an additional functionality.

In order to improve the results of the substitution process, it is important to increase the number of substituable Web services by offering a large range of possibilities. Furthermore, classification may be improved by a more structured organization of the Web services within the communities. Our work follows this line. We propose a network model to represent sets of similar Web services operations. Operations are the nodes and a link account for a similarity relationship between two operations. The main contribution of this work is on the definition and analysis of similarity measures for the functional comparison among Web services. The proposed network model allows the automated identification of substitutable Web services. It differentiates with majority of works efficient mainly for domain-based discovery, but not well suited for the substitution process.

The paper is organized as follows. In section 2, we introduce the similarity network model. We present and comment the similarity functions used in order to cover different situations occurring while searching for a substitutable Web service. In section 3, the presentation of the benchmark used in this study is followed by experimental results on networks topology and components structure of the networks build according to these definitions. Finally, in section 4 the reader will find our conclusion and directions for future work.

## 2   NETWORK MODEL

In this paper, we focus on the functional aspect of semantic Web services. We restrict the definition of a Web service to a set of operations with their input and output parameters. We use the following notations. A Web service is represented by a Greek letter. Each operation labeled by a digit contains a set of input parameters noted I, and a set of output parameters noted O. Each parameter is associated to an ontological concept represented by a letter. Fig. 1 represents a Web service α with two operations 1 and 2, input parameter ontological concepts $I_1 = \{a,b\}$, $I_2 = \{c\}$, and output parameter ontological concepts $O_1 = \{d\}$, $O_2 = \{e,f\}$. In the following, we use for short the word "parameter" rather than "parameter ontological concept" to describe the semantics associated to a parameter.

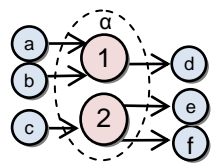

Fig. 1. Schematic representation of a Web service α with two operations 1 and 2 and their parameters.

The similarity network model is based on the similarity between operations. We consider operations rather than Web services as atomic entities for two reasons. First, operations are the entities that are ultimately invoked. Second, it allows getting a more detailed analysis of the similarities. We define a similarity network as a graph whose nodes correspond to operations and links indicate a certain level of similarity between these operations. The nature of the similarity relationship is extremely important. It can be defined in several ways. We propose four operators that reflect different levels of functional similarity. These operators use a semantic matching function to compare the sets of input and output parameters. Thereafter, we describe the four similarity functions that we use to build the networks and we give their interpretation.

### 2.1   Similarity Functions

The four similarity functions are based on the work of [4] and [5] for service discovery. Several operators are presented and used to compare sets of ontological concepts in semantic descriptions. The definition of these operators is made on the key assumption that a user specifies his needs in terms of what he wants to achieve by using a service. In other words, the user knows the goals he wants to get, but the way to reach them is not a major concern. To meet the needs, the request answer can be provided by an individual service or by a set of interacting services. The cornerstones elements are the goals pursued by the user, and they are represented by the output parameters of the service.

We selected four of these operators that reflect different matching situations between user's goals and the outputs provided by the services. We adapted these operator definitions to our goal which is to determine a similarity value between two sets of parameters. The basic idea is that two services are substitutable if they allow to reach the same goal eventually using composite services.

FullSim is defined by analogy with the operator *Match* which reflects the fact that all the user needs are met. *PartialSim* inspired by *Partial match* corresponds to the situation where only a part of the goals is satisfied. Therefore, additional services will be needed to satisfy the request. The two other operators that we selected were introduced in [5] to take into account two situations ignored thus far. *ExcessSim* based on *Excess match* which expresses the case where the published service fully meets the goals of the user and provides more information. *RelationSim* is inspired from *Relation match* which has been introduced for situations where a service can meet the goals but the user cannot provide the inputs to invoke the service. To use such services, additional ones are required.

The similarity functions (*FullSim*, *PartialSim*, *ExcessSim* and *RelationSim)* are defined in terms of set relations. Suppose we want to compare two operations i and j. $I_i$ and $I_j$ are respectively the sets of input parameters of i and j. $O_i$ and $O_j$ are respectively the sets of output parameters of i and j. We hence must compare $I_i$ with $I_j$ and $O_i$ with $O_j$.

*FullSim* means "full similarity". Two operations i and j are fully similar if they offer exactly the same set of output parameters ($O_i = O_j$) and if they have overlapping inputs ($I_i \cap I_j \neq \emptyset$).

*PartialSim* means "partial similarity". Operation j is partially similar to operation i if some output parameters are missing in j ($O_i \supset O_j$) and if the two sets of input parameters overlap ($I_i \cap I_j \neq \emptyset$).

*ExcessSim* means "excess similarity". An operation j is similar with excess to an operation i if j provides all the outputs of i plus additional ones ($O_i \subset O_j$) and if j has at most the inputs of ($I_i \supseteq I_j$).

*RelationSim* means "relational similarity". Two operations i and j have a relational similarity if they have exactly the same outputs ($O_i = O_j$) and if they do not share any common input ($I_i \cap I_j = \emptyset$).

FullSim and PartialSim are symmetric functions, while ExcessSim and RelationSim are asymmetrics.

To achieve the comparison between individual parameters, we take as a basis the classical *exact* and *fail* subsumption relationships introduced in [6]. Let two parameters to be compared. In an *exact* matching, two parameters are similar if they are described by the same ontological concept. The *fail* matching means that there is no subsumption relation between the concepts associated to the parameters.

Each similarity function allows building a specific network. In the following, we use the operators name to refer to the networks obtained with the different similarity functions. FullSim and RelationSim networks, due to the symmetrical nature of the similarity functions, are non-oriented networks. PartialSim and ExcessSim, which are derived from asymmetric functions, are oriented networks.

## 2.2 Interpretation of the Functions

To illustrate the different situations, we show through an example how the similarity functions can be interpreted. Let consider the six operations in Tab. 1.

Tab. 1. Six operations labeled from 1 to 6 with their input and output parameters sets.

|   | Input parameters | Output parameters |
|---|---|---|
| 1 | $I_1$={ZIP} | $O_1$={CITY-NAME} |
| 2 | $I_2$ ={ZIP, GEOGRAPHICALREGION} | $O_2$ ={CITY-NAME} |
| 3 | $I_3$={ZIP} | $O_3$={CITY-NAME, LONGITUDE, LATITUDE} |
| 4 | $I_4$={ZIP} | $O_4$={WEATHERREPORT} |
| 5 | $I_5$ ={CITY-NAME} | $O_5$={WEATHERREPORT} |
| 6 | $I_6$ = {CITY-NAME} | $O_6$ = {WEATHERREPORT, WEATHERREPORTSUBSCR} |

As stated previously, the user goal is the most important aspect to be considered. For example, suppose a user who wants to get the weather report of his city by providing the name and zip code of this city. Searching for operations that satisfy both the inputs and outputs can be too restrictive. Hence, we did not consider this case when designing the similarity functions.

FullSim similarity can be considered as the second best solution, since it includes the expected outputs and some inputs of the request. Operations 4 and 5 are, in this case, two potential candidates. If no operation meets these criteria, the user can relax the constraints on the goal. Suppose that operations 4 and 5 are unavailable, operation 6 which is similar with excess (ExcessSim) to the request, may be the second possibility. It provides a subscription in addition to the weather report. The user may not be interested by this result if he is looking for a free service and if the subscription is a paying service. In another cases, he might be interested in additional outputs such as a list of weather reports for nearby cities, for example.

If the user is always searching for a weather report, but he can only provide a zip code. When operation 4 is unavailable, then no operation can be found using FullSim, PartialSim or ExcessSim similarities. In this case, operation 5 can satisfy the need. It has a relational similarity (RelationSim) with the request because its outputs are identical to the goal, but inputs have nothing in common. This operation cannot be used alone, but it can leads to the goal if it is composed with other operations. In this case, operation 1 can first be invoked and its output parameter, a city name, used to invoke operation 5.

It is important to highlight that the proposed similarity functions have been designed to be complementary. FullSim function is the best solution. Then, PartialSim, ExcessSim and RelationalSim functions can give satisfaction to specific situations that are directly related to the context, as we have seen in the example above.

## 3 STRUCTURE OF THE SIMILARITY NETWORKS

We first present the Web services benchmark used in our experiment. Four similarity networks corresponding to the four similarity functions are built from these data. We provide a global view of the networks and we make a comparison between the similarity networks components and the notion of domain used in Web service classification. Concentrating on the components, we then take a local point of view by studying and comparing the structure of the components from the different networks.

The similarity networks are extracted from SAWSDL-TC [7]. This SAWSDL test collection comes from SemWebCentral, an open source development Web site for the semantic Web. SAWSDL-TC is a service retrieval collection to support the evaluation of the performance of SAWSDL semantic Web service matchmaking algorithms. Although it has not been designed to test Web services substitution models, it best suits our requirements. It is partially composed of real-world Web services that are semantically described. The sets of Web services with similar functionalities are large enough to form reasonable communities. It contains 894 single operation descriptions and 654 are classified into 7 domains. Among them, *economy*, *education* and *travel*, contain more than 80% of the descriptions. *Communication*, *food*, *medical* and *weapon* contain the remaining 20% and their content is relatively uniform. Economy, education, travel and communication are respectively organized into 10, 5, 6 and 2 sub-domains.

### 3.1 Global Structure

The networks exhibit the same structure. A set of small components stand along with isolated nodes. Tab. 2 presents the number of components and proportion of isolated nodes of the networks. Isolated nodes are quite numerous in all the networks. According to these results we can distinguish two types of networks. The first one includes the FullSim, PartialSim and ExcessSim networks while RelationSim is on the second group. Indeed, in the former the networks exhibit similar basic properties. PartialSim presents the lowest proportion of isolated nodes followed by ExcessSim then Fullsim. This behavior is in accordance with the restrictions imposed by the similarity definitions. In other words, there is more Web services that share common output and common inputs than Web services with exactly the same outputs. The differences observed between PartialSim and ExcessSim is problably due to a tougher constraint on the inputs in the later. Similarly, the number of components is quite comparable. Compared to the networks in the first group, RelationSim is quite different. It has at least two times less isolated nodes and the number of components is two times higher. This result suggests that there is a lot of operations that can be used in a substitution process through a composition. It implies to provide other operations to make the connection between the inputs of the desired operation and the inputs of the one extracted from the RelationSim Network.

Tab. 2. Basic properties of the similarity networks.

|  | FullSim | PartialSim | ExcessSim | RelationSim |
|---|---|---|---|---|
| Proportion of isolated nodes | 75% | 57% | 62% | 31% |
| Number of components | 42 | 59 | 66 | 121 |

Note that the distribution of the operations does not reflect the organization of the collection into domains. This structure rather reflects the decomposition of the collection into a reasonable number of sets of similar operations. This is an interesting property. If networks had been composed only of isolated nodes or if we had observed the presence of a giant component, those situations will have lead to an inefficient distribution of the operations. The notion of domain is hence not relevant for substitution.

## 3.2 Local Structure

In order to have a more detailed idea on the influence of the different similarity functions on the networks topology, we investigate and compare the structure of the components. Results show that we can distinguish two types of networks according to the local structure. In FullSim and RelationSim networks, components are organized into cliques. They are dominated by stars in PartialSim and ExcessSim networks. Fig. 2 presents FullSim and ExcessSim networks, without the isolated nodes, that clearly illustrate those two situations.

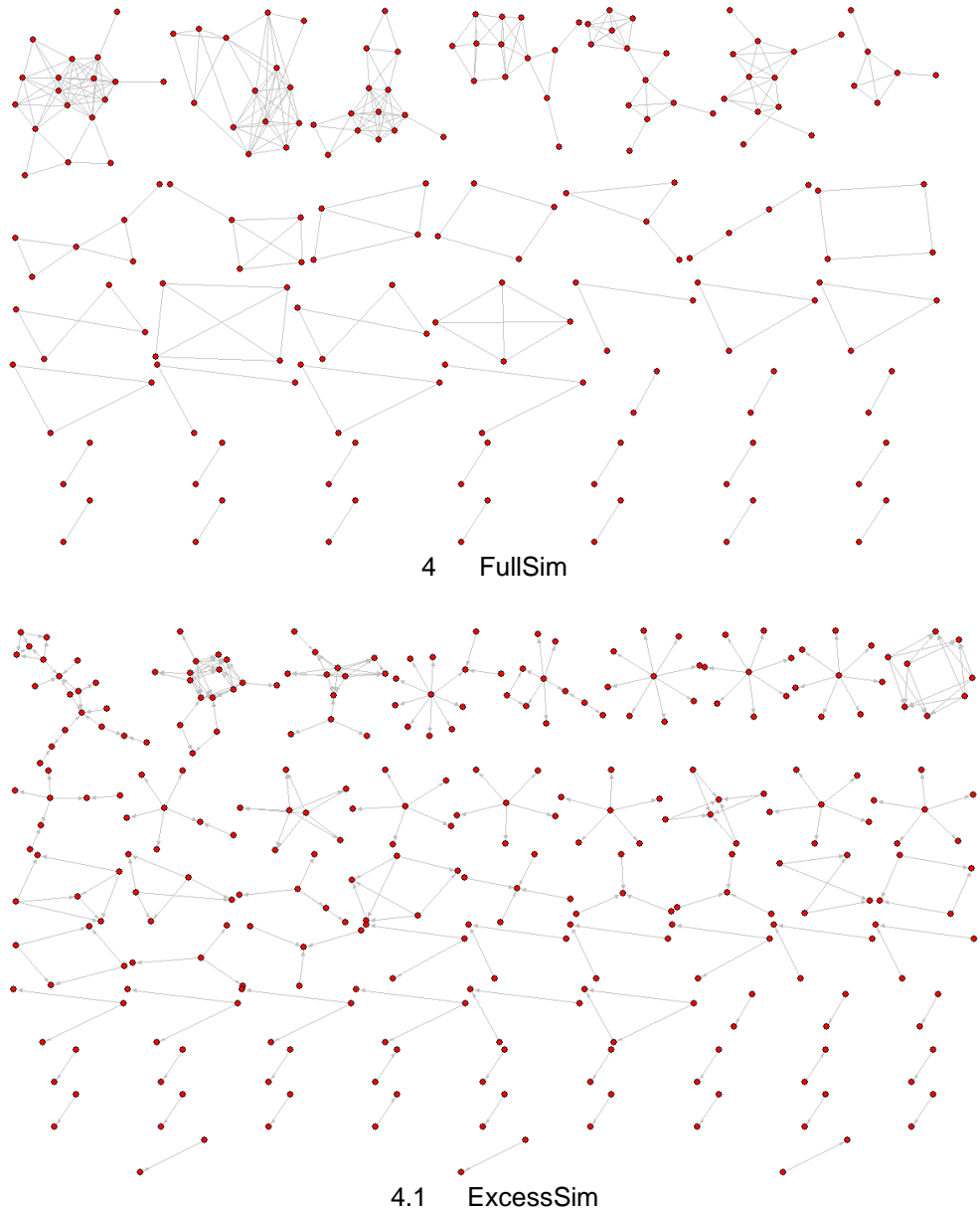

4    FullSim

4.1    ExcessSim

Fig. 2. FullSim and ExcessSim similarity networks where isolated nodes have been discarded.

### A.    Clique as a basic pattern

The clique basic pattern in the FullSim and RelationSim networks give rise to different situations. To illustrate this feature, we present two components of the FullSim network.

The first one is shown on Fig. 3. It contains the get_BOOK and getEBook operations which form a 4-clique. They all produce a single parameter, Book, and they share at least one input parameter. The getEbook operations signatures are identical while getBook and getEbook have one common input parameter (Title). This component includes operations that are all similar according to the FullSim definition. They are all substitutable. If one wants to choose one of them for substitution, they can be distinguishable by their QoS features.

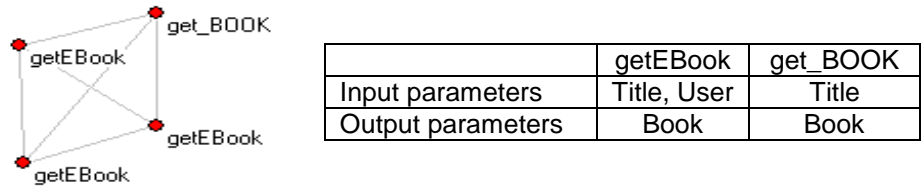

|  | getEBook | get_BOOK |
|---|---|---|
| Input parameters | Title, User | Title |
| Output parameters | Book | Book |

Fig. 3. A 4-clique component of the FullSim network and its operations signature.

The second component on Fig. 4 contains six operations named get_LENDING. They are organized as follow: two 3-cliques and one 2-clique. The six operations have a single and same output parameter (Lending). On the right side of the figure, the links between the six operations are labeled by the concept of the input parameters shared by two adjacent operations. Unlike in the previous case, operations in this component are not all similar according to the Fullsim definition. Only operations that are within a clique follow this definition. This component includes three sets of similar operations which are organized into cliques. Two operations that are not in the same clique are similar according to the RelationSim definition. They have common outputs but their inputs do not overlap. Hence, when searching for substitutable operations in a FullSim component, all the cliques must be considered independently.

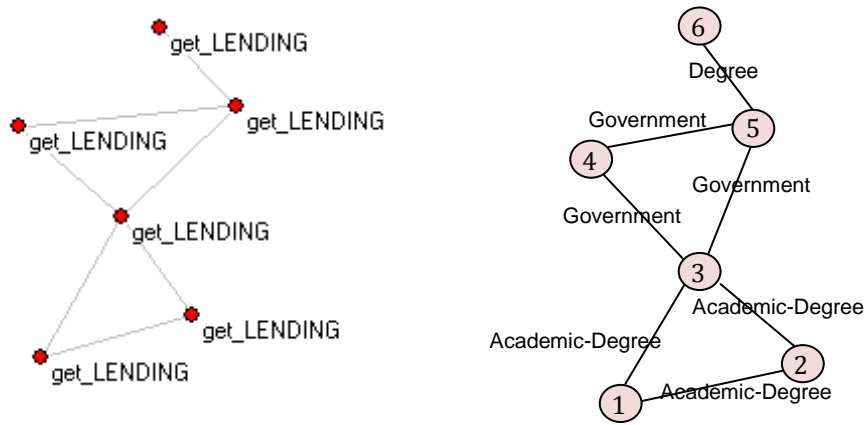

Fig. 4. A component of the FullSim network with 3 cliques (links labels are common inputs of two adjacent operations).

Distribution of similar operations within components allows classifying them according to their functionality. A component is a set of operations that have identical outputs. Operations with at least one common input are grouped within cliques in this component. A component is not a monolithic block of similar operations. It can be decomposed into a set of communities characterized by a clique. The clique pattern in the components allows a finer characterization of the notion of operations community.

Note that in the RelationSim network, the clique organization is more pronounced than in the FullSim network and some big components form a complete graph.

### B. Star as a basic pattern

The structure of the components in PartialSim and ExcessSim networks clearly differs from the previous ones. Whereas FullSim and RelationSim components are clique-like, PartialSim and ExcessSim components are rather organized as stars. We consider the PartialSim network in order to illustrate the two most typical situations observed in these cases.

Fig. 5 shows a star component of the PartialSim network. The get_FILM central operation produces less output parameters compared to peripheral operations. It produces only the parameter Film while the five others produce additional ones. It also has common input parameters with the peripheral operations. In this case, these six operations share a unique parameter (Title). The peripheral operations have a unique and same potential substitute. This substitute, which is pointed by the others, can replace them being aware that it provides less output that may be desired.

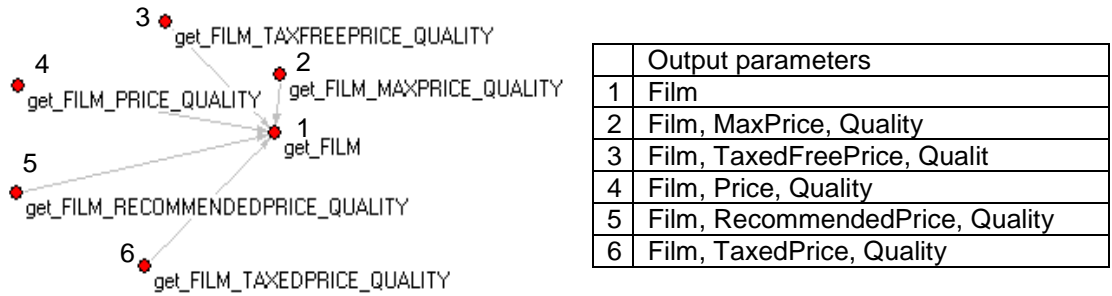

|   | Output parameters |
|---|---|
| 1 | Film |
| 2 | Film, MaxPrice, Quality |
| 3 | Film, TaxedFreePrice, Qualit |
| 4 | Film, Price, Quality |
| 5 | Film, RecommendedPrice, Quality |
| 6 | Film, TaxedPrice, Quality |

Fig. 5. A 6-nodes star-like component of the PartialSim network and its operations output parameters.

Fig. 6 shows a component with 15 nodes made up with nested stars. Four operations may be replaced by others: get_DESTINATION_HOTEL, get_ACTIVITY_HOTEL, and two get_SPORTS_HOTEL operations. Unlike in the previous example, this component contains several substitutes. Some may replace only one operation; the get_HOTEL at the left end side is only a substitute for get_ACTIVITY_HOTEL. Others may replace several operations; all the get_HOTEL operations of the right end side can be substitutes for get_DESTINATION_HOTEL and the two get_SPORTS_HOTEL operations. Like for the clique organization, the concept of similarity can be refined within a component. It occurs when the component does not simply contain a simple basic pattern, but a more complex structure built from this pattern.

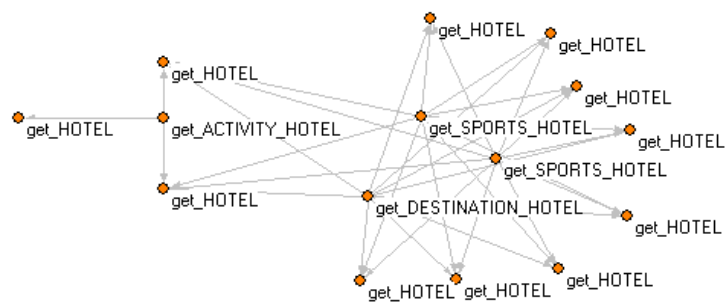

Fig. 6. A 15-nodes component of the PartialSim network with nested stars.

When observing the same components in the ExcessSim network, they differ in two points. The links are oppositely oriented and some of them disappear if one operation has more input parameters than the one to which it is compared.

## 5  CONCLUSION

In this paper, we analyze the structure of similarity networks extracted from SAWSDL-TC1. The networks are built from a model that represents similarity relationships between Web services operations functionalities. Two operations are similar if they share common features regarding their input and output parameter sets. We defined a set of functions that represent different degrees of similarity between operations. We compared the structure of the networks obtained with the different similarity functions. This comparative study shows that the networks share the same global structure. They are characterized by a large number of isolated nodes. It evolves from 30% to 75% depending on the more or less restrictive definition of similarity function. The remaining nodes are organized into a number of small components of similar operations. From the components analysis, we identified two classes of networks. In FullSim and RelationSim networks, the organizational basic pattern is the clique while it is a star in PartialSim and ExcessSim networks.

In clique structured networks, a component is a clique or a set of cliques. In a FullSim component, all the operations of a clique have identical output parameters and their input parameters overlap. Two operations that do not belong to the same clique have disjoint input parameters. In the RelationSim network, components are strongly connected and can be complete graphs.

In the star structured networks, a component is a star or a set of stars. In a PartialSim component, operations that are pointed by a link have less output parameters than the operations pointing toward them. Two linked operations have overlapping input parameters. Two operations that do not belong to the same star have disjoint input parameters. In an ExcessSim component, links are oppositely directed compared to the same component in the PartialSim network. Additionaly, some links disappear because of the restriction on the input parameters sets.

This proposed classification reveals two levels of similarity. The first one is the component and the second one is the component basic pattern (clique or star). A basic pattern within a component is a community that groups a set of similar and substitutable operations.

This analysis allowed an accurate understanding of the functional similarity relationships between Web services operations. We are extending this work by using the similarity networks as a structure to discover and substitute Web services during the composition process.